\documentstyle[12pt,amsfonts,amssymb]{report}
\begin{document}
\begin{titlepage}
\pagestyle{empty}
\title{Gelfand-Fuchs cohomology and M-Theory as a Topological Quantum Field Theory}
\author{Ioannis P. \ ZOIS\thanks{izois\,@\,maths.ox.ac.uk and
izois\,@\,cc.uoa.gr}\\
\\
Mathematical Institute, 24-29 St. Giles', Oxford OX1 3LB\\}
\date{}
\maketitle
\begin{abstract}

We propose a Lagrangian density for M-Theory which is purely topological
using \emph{Gelfand-Fuchs cohomology} which characterises up to homotopy
$\Gamma _q $-structures and hence foliations in particular. Then
using S-duality, we conjecture on the existence of certain
plane fields on $S^{11}$. Finally we calculate the partition function of 
the theory.\\



To my brother Demetrios and to E.\\

\end{abstract}
\end{titlepage}

\section{Short review of current status in M-Theory}

We shall start with a brief overview of "old" superstring theory, namely
the picture arround the end of '80's before advances in dualities and
p-branes.

What we actually had was not one but five consisten superstring theories
in D=10, namely types I, IIA, IIB, heterotic SO(32) and heterotic $E_8
\times E_8 $ which was an embarrassment of richness. These theories were
related via a pertutbative symmetry called T-duality when compactified in
D=9 (this duality is a symmetry of theories with compactified spatial
dimmensions where in the case of Calabi-Yau 3-folds T-duality takes the
form of
mirror symmetry). Moreover we believed that D=10 N=1 supergravity
(various versions of it) was the
low energy limit (or infinite string tension limit) of these superstring
theories in D=10.

There was however another piece of knowledge available which was somehow
"forgotten", and that was D=11 N=1 supergravity. This theory emerged as
folows: assuming that spacetime carries a metric with Minkowski signature,
namely that there is only one timelike coordinate and assuming also that
there are zero modes with no grater than 2 spin, then supersymmetry puts
restrictions on dimensionality of spacetime, hence $D_{max} = 11$. This
made phenomenologists over the years to study all possible representations
of super-Poincare in various dimensions up to D=11. One of the
possibilities then was this maximal D=11 N=1 supergravity. However this
theory was rather unpopular because it was overshadowed by superstring
theory in D=10 (with which no apparent relation was known) but also
perhaps for a more important reason: it was proved to be
non-renormalisable. At this point we make a prothysteron: this defect is
not possible to be overcome and one reminicent of it is the fact that the
supermembrane in D=11 has continoum spectrum; there is however an attempt
to encorporate this feature in the framework of matrix models.

Situation changed rapidly around mid '90's (or perhaps earlier), the most
important advances being:

1.) It was proved that various versions of D=10 N=1 supergravity admit
strings as solutions, a fact that now puts the two theories on a rather
equal footing.

2.)  New evidence for a type of non-perturbative symmetry called S-duality
was found. This symmetry interchanges the strong/weak coupling regions of
a theory as well as elementary/soliton solutions (and hence topology and
dynamics; it is essentially a Hodge star duality between field strengths).
This S-duality emerged as a generalisation of electric/magnetic duality in
Maxwell theory and as furter evidence of Montonen-Olive duality in
supersymmetric Yang-Mills theories. Now eventually S-duality has lead to
string/5-brane duality in D=10 and to membrane/5-brane duality in D=11
(let us restrict ourselves to the old brane-scan for the moment and forget
D-branes also).

3.) The observation that the five different superstring theories in D=10
was an artifact of perturbation theory; these theories should emerge
from one non-perturbative theory in D=11 which was called M-Theory.

4.) The relation between supermembranes in D=11 with D=11 N=1 supergravity
as well as with D=10 superstring theories; the later is done by a process
which now comes under the name \emph{"dimensional reduction process"}.
This is a general scheme which has been generalised to incorporate
arbitrary dimension D and arbitrary p-branes.

5.) Applying S-duality in D=11 supermembranes we get the membrane/5-brane
duality in D=11, something which we mentioned earlier.

(see references \cite{duff}, \cite{duff1}, \cite{town}, \cite{wit},
\cite{west}, \cite{witten}, \cite{mo} and referencees therein).

\section{A topological Lagrangian for M-Theory}

Hence we have something called M-Theory consisting of membranes and
5-branes living on an 11-manifold which is non perturbative. This theory
has a very intriguing feature: we can only extract information about it
from its limiting theories, namely either from D=11 N=1 supergravity or
from superstrings in D=10. This is so because this theory is
\emph{genuinly} non
perturbative for a reason which lies in the heart of manifold topology:

Let us recall that in string theory, the path integral involves summation
over all topologically distinct diagrams (same for point
particles of
course). Strings are 1-branes hence in time they swep out a 2-manifold. At
the tree level then we need all topologically distinct simply connected
2-manifolds (actually there is only one, as topology tells us) and for
loop corrections, topology again says that topologically distinct non
simply connected 2-manifolds are classified by their genus, so we sum up
over all Riemann surfaces with different genus.

It is clear then that for a perturbative quantum field theory involving
p-branes we have to sum upon all topologically distinct (p+1)-dim
manifolds: simply connected ones for tree level and non simply connected
ones for loop corrections. Thus \emph{we must know before hand the
topological
classification of manifolds} in the dimension of interest. That is the
main problem of manifold topology in mathematics.

But now we face a deep and intractable problem: geometry tells us,
essentially via a no-go theorem which is due to Whitehead from late '40's,
that:
\emph{"we cannot classify non simply connected manifolds with dimension
greater or equal to 4"}! Hence for p-branes with p greater or equal to 3,
all we can do via perturbative methods is up to tree level!

What happens for 3-manifolds then (hence for membranes)?
The answer from mathematics is that we \emph{do not know} if all
3-manifolds can be
classified! So even for 2-branes it is still unclear whether perturbative
methods work (up to all levels of perturbation theory)!\\

The outlet from this situation that we propose here is not merely to look
only at non perturbative aspects of these theories, as was done up to now,
but to abandon perturbative methods completely from the very beginning.
There is only one way known up to now which can achieve this "radical"
solution to our problem: \emph{formulate
the theory as a Topological Quantum Field Theory} and hence get rid of all
perturbations once and for all.\\

Let us explain how this can be acieved.\\

Our approach is based on one physical \emph{"principle"}:\\

\textsl{A theory containing \emph{p-branes} should be formulated on an
m-dim manifold
which \emph{admits $\Gamma _q$-structures}, where $q=m-p-1$}.\\

{\bf N.B.} 

Although we used in our physical principle $\Gamma _q$-structures
which are more general than foliations, we shall use both these terms
meaning essentially the same structure. The interested reader may
refer to \cite{lawson} for example to see the precise definitions which are
quite complicated. The key point however is that the difference
between $\Gamma _q$-structures (or Haefliger structures as they are
most commonly known in topology) and codim-q foliations is essentially the
difference between \textsl{transverse} and \emph{normal}. This does
not affect any of what we have to say, since Bott-Haefliger theory
of characteristic classes is formulated for the most general case, namely
$\Gamma $-structures. We would also like to mention the relation
between $\Gamma $-structures and $\Omega $-spectra which is currently
an active field in topology.\\

(For D-branes we need a variant of the above principle, namely we need
what are called \emph{plane foliations} but we shall not elaborate on this
point here).\\

One way of thinking about this principle is that it is analogous to
the ``past histories'' approach of quantum mechanics. Clearly in
quantum level one should integrate over all foliations of a given codim.\\

\textsl{A piece of warning here:} this principle does not imply that \emph{all}
physical process between branes \emph{are} described by foliations. Although
the group of foliations is huge, in fact comparable in size with the
group of local diffeomorphisms \cite{william}, and foliations can be really ``very
nasty'', we would not like to
make such a strong statement. What is definitely true though is that
\emph{some} physical process \emph{are indeed} described by foliations, hence
\textsl{at least} this condition \emph{must} be satisfied because of them.\\

{\bf Note:}

Before going on furter, we would like to make one crucial remark: this
principle puts severe restrictions on the topology that the underlying
manifold may have, in case of M-Theory this is an 11-manifold. It is also
very important if the manifold is \emph{open} or \emph{closed}. This may
be of some help, as we hope, for the compactification problem of string
theory or even M-Theory, namely how we go from D=10 (or D=11) to D=4 which
is our
intuitive dimension
of spacetime. We shall address this question in the next section. The final comment is this: this principle puts
\emph{absolutely no restriction} to the usual quantum field theory for
point particles in D=4, e.g. electroweak theory or QCD. This is so because in
this case spacetime is just ${\bf R^4}$ which is non compact and we have
0-branes (point particles) and consequently 1-dim foliations for which the
integrability condition is trivially satisfied (essentially this is due
to a deep result of Gromov for foliations on open manifolds, which
states that all open manifolds admit codim 1 foliations; in striking contrast,
closed  manifolds admit codim 1 foliations iff their Euler
characteristic is zero, see
for example in \cite{lawson}, \cite{gromov} or references therein).\\

If we believe this principle, then the story goes on as follows: we are on
an
11-manifold, call it M for brevity and we want to describe a theory
containing 5-branes for example (and get membranes from S-duality). Then M
should admit 6-dim foliations or equivalently codim 5 foliations. We know
from Haefliger that the $\Gamma _q $-functor, namely the functor of codim
q Haefliger structures and in particular codim q foliations, is
representable. Practically this means that we can have an analogue of
Chern-Weil theory which characterises foliations of M up to homotopy using
cohomology classes of M. (One brief comment for foliations: one way
of describing Haefliger
structures more generally is to say that they \emph{generalise fibre
bundles in exactly
the same way that fibre
bundles generalise Cartesian product}. This observation is also
important when mentioning \textsl{gerbes} later on).\\

In fact it is proved that the correct cohomology to classify Haefliger
structures up to homotopy (and hence foliations which constitute a
particular example of Haefliger structures) is the \emph{Gelfand-Fuchs}
cohomology. This is a result of Bott and Haefliger, essentially
generalising an earlier result due to Godbillon and Vey which was dealing
only with codim 1 foliations, \cite{gv}.\\

Now we have a happy coincidence: the Bott-Haefliger class for a codim 5
foliation (which, recall, is what we want for 5-branes on an 11-manifold)
is exactly an 11-form, something that fits well with using it as a
Lagrangian density!

The construction for arbitrary codim q foliations goes as follows: let
$F$ be a codim q foliation on an m-manifold M and suppose its
normal
bundle $\nu (F)$ is orientable. Then $F$ is defined by a
global decomposable q-form $\Omega $. Let $\{(U_{i},X_i)\}_{i\in
I}$ be a locally finite cover of distinguished coordinate charts on M with
a smooth partition of unity $\{\rho _i\}$. Then set

$$ \Omega = \sum _{i\in I} \rho _{i}dx_{i}^{m-q+1}\wedge ... \wedge
dx_{i}^{m} $$

Since $\Omega $ is integrable,

\begin{equation}
d\Omega =\theta \wedge \Omega ,
\label{eq1}
\end{equation}

where $\theta $ \emph{some} 1-form on M. The (2q+1)-form

\begin{equation}
\gamma = \theta \wedge (d\theta )^{q},
\label{eq2}
\end{equation}

is closed and its de Rham cohomology class is independent of all choices
involved in defining it, depending only on homotopy type of $F$.
That's the class we want.

Clearly for our case we are on an 11-manifold dealing with 5-branes, hence
6-dim foliations, hence codim 5 and thus the class $\gamma $ is an
11-form.

This construction can be generalised to arbitrary $\Gamma
^{r}_{q}$-structures as a mixed de Rham-Cech cohomology class and thus
gives an element in $H^{2q+1}(B\Gamma ^{r}_{q};{\bf R})$, where $B\Gamma
^{r}_{q}$ is the classifying space for $\Gamma ^{r}_{q}$-structures.
Note that in fact the BHGV class is a cobordism invariant of codim q foliations
of compact (2q+1)-dim manifolds. This construction gives \emph{one}
computable characteristic class for foliations. Optimally we would
like a generalisation of the Chern-Weil construction for $GL_q$. That
is we would like an abstract GDA with the property that for any codim
q foliation $F$ on a manifold M there is a GDA homomorphism
into the de Rham algebra on M, defined in terms of $F$ such
that the induced map on cohomology factors through a universal map
into $H^*(B\Gamma _{q}^{r};{\bf R})$. This algebra is nothing more
than the \emph{Gelfand-Fuchs} Lie coalgebra of \emph{formal} vector fields in
one variable. 

More concretely, let $\Gamma $ be a transitive Lie-pseudogroup acting
on ${\bf R^n}$ and let $a(\Gamma )$ denote the \textsl{Lie algebra of
formal $\Gamma $ vector fields} associated to $\Gamma $. Here a vector
field defined on on $U\subset {\bf R^n}$ is called a $\Gamma $ vector
field if the local 1-parameter group which it engenders is $\Gamma
$ and $a(\Gamma )$ is defined as the inverse limit
$$a(\Gamma )=lim_{\leftarrow}a^{k}(\Gamma )$$
of the k-jets at 0 of $\Gamma $ vector fields. In the pseudogroup
$\Gamma $ let $\Gamma _0 $ be the set of elements of $\Gamma $ keeping
0 fixed and set $\Gamma _{0}^{k} $ equal to the k-jets of elements in
$\Gamma _0 $.\\

Then the $\Gamma _{0}^{k} $ form an inverse system of Lie groups and
we can find a subgroup $K\subset lim_{\leftarrow}\Gamma _{0}^{k}$
whose projection on every $\Gamma _{0}^{k} $ is a maximal compact
subgroup for $k>0$. This follows from the fact that the kernel of the
projection $\Gamma _{0}^{k+1}\rightarrow \Gamma _{0}^{k}$ is a vector
space for $k>0$. The subgroup K is unique up to conjugation and its Lie
algebra k can be identified with a subalgebra of $a(\Gamma )$.

For our purposes we need the cohomology of basic elements rel K in
$a(\Gamma )$, namely $H(a(\Gamma );K)$ which is defined as follows:
Let $A\{a^k(\Gamma )\}$ denote the algebra of multilinear alternating
forms on $a^k(\Gamma )$ and let $A\{a(\Gamma )\}$ be the direct limit
of the $A\{a^k(\Gamma )\}$. The bracket in $a(\Gamma )$ induces a
differential on $A\{a(\Gamma )\}$ and we write $H\{a(\Gamma )\}$ for
the resulting cohomology group. The relative group $H^{*}(a(\Gamma
);K)$ is now defined as the cohomology of the subcomplex of
$A\{a(\Gamma )\}$ consisting of elements which are invariant under the
natural action of K and annihilated by all inner products with
elements of k. Then the result is:

\textsl{Let $F$ be a $\Gamma $-foliation on M. There is an
algebra homomorphism}

$$\phi :H\{a(\Gamma );K\}\rightarrow H(M;{\bf R})$$

which is a natural transformation on the category $C(\Gamma )$.

The construction of $\phi $ is as follows:

Let $P^k(\Gamma )$ be the differential bundle of k-jets at the origin
of elements of $\Gamma $. It is a principal $\Gamma
_{0}^{k}$-bundle. On the other hand $\Gamma $ acts transitively on the
left on $P^k(\Gamma )$. Denote by $A(P^{\infty}(\Gamma ))$ the direct
limit of the algebras $A(P^{k}(\Gamma ))$ of differential forms on
$P^k(\Gamma )$. The invariant forms wrt the action of $\Gamma $
constitute a differential subalgebra denoted $A_{\Gamma }$. One can
then prove that it is actually isomorphic to $A(a(\Gamma ))$.

Now let F be a foliation on M and let $P^k(F)$ be the differentiable
bundle over M whose fibre at every point say $x\in M$ is the space of k-jets at
this point of local projections that vanish on $x$. This is a $\Gamma
_0^k $-principal bundle. Its restriction is isomorphic to the inverse
image of the bundle $P^k(\Gamma )$, hence the differential algebra of
$\Gamma $-invariant forms on $P^k(\Gamma )$ is mapped in the algebra
$A(P^k(F))$ of differential forms on $P^k(F)$. If we denote by
$A(P^{\infty}(F))$ the direct limit of $A(P^{k}(F))$ we get an
injective homomorphism $\phi $ of $A(a(\Gamma ))$ in $A(P^{\infty}(F
))$ commuting with the differential.

This homomorphism is compatible with the action of K, hence induces a
homomorphism on the subalgebra of K-basic elements. But the algebra
$A(P^{k}(F );K)$ of K-basic elements in $A(P^{k}(F))$ is isomorphic to
the algebra of differential forms on $P^k(F)/K$ which is a bundle over
M with contractible fibre $\Gamma _0^k/K$.Hence $H(A(P^k(F);K))$ is
isomorphic via the de Rham theorem to $H(M;{\bf R})$. The homomorphism
$\phi $ is therefore obtained as the composition

$$H(a(\Gamma );K)\rightarrow H(A(P^{\infty }(F);K))=H(M;{\bf R})$$

But we think that is enough with \emph{abstract nonsense}
formalism. Let us make our discussion more \textsl{down to earth}:

Consider the GDA (over {\bf R})

$$WO_{q}=\wedge (u_1, u_3,..., u_{2(q/2)-1})\otimes P_q(c_1,...,
c_q)$$

with $du_i=c_i $ for odd i and $dc_{i}=0$ for all i and
$$W_{q}=\wedge (u_{1},u_{2},...,u_{q})\otimes P_{q}(c_1,...,c_q)$$

with $du_{i}=c_{i}$ and $dc_{i}=0$ for i=1,...,q where $deg
u_{i}=2i-1$, $deg c_{i}=2i$ and $\wedge $ denotes exterior algebra,
$P_q $ denotes the polynomial algebra in the $c_{i}$'s mod elements of
total degree greater than 2q. The cohomology of $W_q $ is the Gelfand
Fuchs cohomology of the Lie algebra of formal vector fields in q
variables. We note that the ring structure at the cohomology level is
trivial, that is all cup products are zero. Then the main result is
that there are homomorphisms
$$\phi :H^{*}(WO_{q})\rightarrow H^{*}(B\Gamma ^{r}_{q};{\bf R})$$

$$\tilde {\phi }:H^{*}(W_q)\rightarrow H^{*}(\tilde{B\Gamma _{q}^{r}};{\bf
R})$$

for $r \geq 2$ with the following property ($\tilde{B\Gamma _{q}^{r}}$
denotes the classifying space for \emph{framed} foliations): If $F$ is a codim
q $C^{r}$ foliation of a manifold M, there is a GDA homomorphism
$$\phi _{F}:WO_{q}\rightarrow \wedge ^{*}(M)$$
into the de Rham algebra on M, defined in terms of the differential
geometry of $F$ and unique up to chain homotopy, such that on
cohomology we have $\phi _{F}=f^{*}\circ \phi $, where
$f:M \rightarrow B\Gamma ^{r}_{q}$ classifies $F$. If the
normal bundle of $F$ is trivial, there is a homomorphism
$$\tilde {\phi _{F}}:W_q \rightarrow \wedge ^{*}(M)$$
with analogous properties.\\

Combining this result with the fact that $B\tilde{\Gamma ^{0}_{q}}$ is
\emph{contractible}, we deduce that a foliation is essentially
determined by the structure of its normal bundle; the \emph{Chern}
classes of the normal bundle are contained in the image of the map
$\phi $ above but we have \emph{additional} non trivial classes in the
case of foliations (which are rather difficult to find though), one of
which is this BHGV class which we
constructed explicitly and it is the class we use as a Lagrangian
density which is purely topological since its degree fits nicely for
describing 5-branes.

There is an alternative approach due to Simons \cite{simons} which avoids passing to
the normal bundle using circle coefficients.  What he actually does is
to associate to a principal bundle with connection a family of
characteristic homomorphisms from the integral cycles on a manifold to
$S^{1}$ and then defining an extension denoted $K^{2k}_{q}$ of
$H^{2k}(BGL_{q};{\bf Z})$. This approach is related
to \emph{gerbes}. A gerbe over a manifold is a construction which
locally looks like the Cartesian product of the manifold with a line
bundle. Clearly it is a special case of foliations (remember our
previous comment on foliations). However this
approach actually suggests that they might be equivalent, if the
approach of Bott-Haefliger is equivalent to that of Simons,
something which is not known.\\

\textsl{Now the conjecture is that the \emph{partition function} of this
Lagrangian is related to the invariant introduced in \cite{z}.}\\

In order to establish relation with physics, we must make some
identifications. The 1-form $\theta $ appearing in the Lagrangian has
no direct physical meaning. In physics it is assumed that a 5-brane
gives rise to a 6-form gauge field denoted $A_6$ whose field strength
is simply
\begin{equation}
dA_{6}=F_{7}
\label{eq3}
\end{equation}
The only way we can explain geometrically this is that this 6-form is
the Poincare dual of the 6-chain that the 5-brane sweps out as it
moves in time. 

We know that since we have S-duality between membranes and 5-branes,
in an obvious notation one has
\begin{equation}
F_{7}=*F_{4}
\label{eq4}
\end{equation}
which is the S-duality relation, where
\begin{equation}
F_{4}=dA_{3}
\label{eq5}
\end{equation}

Observe now that the starting point for 5-brane theory is $A_6$ where
the starting point to construct the BHGV class was the 5-form $\Omega
$. How are they related?

There are three obvious possibilities:

I. $d\Omega =A_6$
That would imply that $A_6 $ is pure gauge.

II.$dF_{4}=\Omega $
This is trivial because it implies $d\Omega =0$, hence $d\Omega
=\theta \wedge \Omega =0$.

III. The only remaining possibility is
\begin{equation}
*A_{6}=\Omega
\label{eq6}
\end{equation}

We call this \emph{``reality condition''}. So now in principle we can substitute
equations (6) and (1) into (2) and get an expression for the Lagrangian
which involves the gauge field $A_6$.

The Euler-Lagrange equations which are actually analogous to D=11 N=1
supergravity Euler-Lagrange equations (see equation 8 below) read:
\begin{equation}
d*d\theta + \frac{1}{5}(d\theta )^{5}=0
\label{eq7}
\end{equation}

The on-shell relation with D=11 N=1 supergravity is established as
follows: recall that the bosonic sector of this supergravity theory is
$$\int F_{4}\wedge F_{4}\wedge A_{3}$$
 where $F_{4}=dA_{3}$ with
Euler-Lagrange equations
\begin{equation}
d*F_{4}+\frac{1}{2}F_{4}\wedge F_{4}=0
\label{eq8}
\end{equation}

Constraining $A_3 $ via (8), by (5), (4), (3), (6) and (1) we get a constraint
for $\theta $ which can be added to the class $\gamma $ as a Lagrange
multiplier.\\

 Let us note that a Lagrangian density with only the BHGV class would give
as Euler-Lagrange equations the condition that $\theta $ is closed.\\

In principle, one must end up with an equivallent theory starting with
membranes (that's due to S-duality), provided of course a suitable
class was found. Clearly the BHGV class for a membrane would be a
\emph{17-form}.

The final comment is this: it is a rather intriguing feature of this
approach that although we start with 5-branes which give rise to
\emph{6-forms}
as gauge potentials, we end up with a characteristic class for foliations
as a Lagrangian density which  involves \emph{1-forms}. If we see this
1-form $\theta $ appearing in the Lagrangian as a gauge potential, that
would imply the existence of \emph{point particles} which are really the
fundamental elements. This idea is in accordance to what was described in
\cite{banks} for instance, since matrix models use point particles'
degrees of freedom.

\section{Plane fields}

We now pass on to the second question raised in this work, namely the
restrictions on the topology of the underlying manifold of a theory
containing p-branes via our physical principle.

It is clear from the definition that the existence of a 
foliation  of certain dim, say d (or equivalently codim q=n-d) on an
n-manifold (closed) depends:

a.) On the existence of a dim d subbundle of the tangent bundle

b.) On this d-dim subbundle being integrable.

The second question has been answered almost completely by Bott and in
a more general framework by
Thurston. Bott's result dictates that for a codim q subbundle of the
tangent bundle to be integrable, the ring of Pontrjagin classes of the
subbundle with degree $>2q$ must be zero. There is a secondary
obstruction due to Shulman involving certain Massey triple products
but we shall not elaborate on this. However Bott's result suggests
nothing for question a.) above. Let us also mention that this result of
Bott can be deduced by another theorem due to Thurston which states
that the classifying space $B\tilde{\Gamma ^{\infty }_{q}}$ of smooth
codim q framed foliations is (q+1)-connected. 

On the contrary, Thurston's result reduces the existence
of codim $q>1$ foliations (at least up to homotopy) to the existence of
\emph{q-plane fields}. This is a deep question in differential
topology, related to the problem of classification of closed manifolds
according to their \textsl{rank}.

Now the problem of existence of q-plane fields has been answered
\emph{only for some cases for spheres $S^n$ for various values of
n,q} \cite{steenrod}.
In particular we know everything for spheres of dimension 10 and
less. We should however mention a theorem due to Winkelnkemper
\cite{win} which
is quite general in nature and talks about simply connected compact
manifolds of dim n greater than 5. If n is not 0 mod 4 then it admits
a so-called \emph{Alexander decomposition} which under special
assumptions can give a particular
kind of a codim 1 foliation with $S^{1}$ as space of leaves and a
surjection from the manifold to $S{1}$. If n is 0 mod 4 then
the manifold admits an
Alexander decomposition iff its signature is zero.

Let us return to string theory now: String theory works in D=10 and in this case we have the old brane-scan
suggesting the string/5-brane duality. The new brane-scan contains all
p-branes for $p\leq 6$ and some D-7 and 8-branes are thought to
exist. However topology says that for a sphere in dim 10 we can have
only dim 0 and dim 10 plane fields (in fact this is true for all even
dim spheres), hence by Thurston only dim 0 and
dim 10 foliations and then our physical principle suggests that $S^{10}$
is ruled out as a possible underlying topological space for string theory.\\

What about M-Theory in D=11 then?\\

For the case of $S^{11}$ then it is known that $S^{11}$ admits a 3-plane field, hence by our physical
principle a theory containing membranes \emph{can} be formulated on
$S^{11}$. For $S^{11}$ nothing is known for the existence of q-plane
fields for q greater than 3. But now we apply S-duality between membranes/5-branes
 and conjecture that:\\

\emph{$S^{11}$ should admit 5-plane fields}.\\

Let us close this section with two  remarks: 

1. There is extensive work in
foliations with numerous results which actually insert many extra
parameters into their study, for example metric aspects, existence of
foliations with compact leaves (all or at least one or exactly one),
with leaves diffeomorphic to ${\bf R^n}$ for some n etc. We do not have
a clear picture for the moment concerning imposing these in physics.
Let us only mention
one particularly strong result due to Wall generalising a result of
Reeb \cite{reeb}: if a closed
n-manifold admits a codim 1 foliation whose leaves are homeomorphic to
${\bf R^{n-1}}$, then by Thurston we know that its Euler characteristic
must vanish, but in fact we have more: it has to be the n-torus!\\

The interesting point
however is that although all these extended objects theories in
physics are expressed as $\sigma $ models \cite{z}, hence they
involve metrics on the manifold (target space) and on the worldvolumes ie on the
leaves, in our approach the metric is only used in the reality
condition (6) which makes connection with physical fields (that is
some metric on the target space) where at the same time we do not use any
metric on the source space (worldvolumes-leaves of the foliation).\\

2. In \cite{z} another Lagrangian density was proposed. It is
different from the one described here but they are related in an
analogous way to the relation between the Polyakov and Nambu-Goto (in
fact Dirac \cite{dirac}) actions for the free bosonic string: extended
objects basically immitate string theory and we have two formalisms:
the $\sigma $ model one which is the Lagrangian exhibited in \cite{z}
using Polyakov's picture of $\sigma $ models as flat principal bundles
with structure group the isometries of the metric on the target space
\cite{polyakov}; yet we also have the \emph{embedded surface} picture
which is the Dirac (Nambu-Goto) action and whose analogue is described
in this work.

\section{The partition function}

Let us try to see what we can say about the \emph{quantum theory} with the
BHGV
class as Lagrangian density. Hopefully we can actually say rather a lot.

The starting point will be the crucial observation that the original
Godbillon-Vey class for codim 1 foliations is simply
$$AdA$$,
which is actually the \emph{"Abelian part"} of the 3-dim Chern-Simmons
form. This "topological term" was for the first time used in physics to
describe some peculiar properties of the spin and statistics of
\emph{skyrmions}, see \cite{zee}. At that moment of course it was
\emph{not} realised
that this was actually a characteristic class for foliations. Skyrmions
are soliton solutions of 3-dim non-linear $\sigma $ model with target the 
group $O(3)$ and source, in most cases ${\bf R^3}$ compactified to $S^3$.
At
that paper it was described as the \emph{Hopf invariant}, emerging from
the
first known Hopf fibration $S^2 \rightarrow S^3$. This class is also a
characteristic class for flat foliations of bundles. The picture is
consistent because from Polyakov we know that non-linear $\sigma $ models
can be seen as flat bundles (for more details on this see \cite{z}).

The Hopf invariant is related to linking number between curves in ${\bf
R^3}$. The non-abelian 3-dim generalisation was related to knots and the
Jones' polynomial in \cite{witten}. One of the main differences in these
two cases is the "Dirac quantisation condition" of the parameter needed in
the non-abelian case.

This observation make as to believe that the answer to the problem of
quantising the BHGV class in 11-dim is actually reduced to quantisation of
Abelian Chern-Simons theory. The answer was given by Schwarz \cite{schw}:
the result
is the \emph{"Ray-Singer analytic torsion"} of the Laplacian on our
11-manifold. As
far as we know it is not yet proved the long standing conjecture that
this actually coincides with the
combinatorial Reidemeister torsion of algebraic topology.\\

Let us make some remarks on this point: Schwarz's result is actually more
general than what we need in this particular case. He uses the notion of
\emph{elliptic resolvent} of a linear functional. In the special case
where this elliptic resolvent is actually an "ellitpic complex", he proves
that the partition function of the functional is actually the analytic
torsion of the corresponding elliptic complex. In the even more special
case where the functional is actually of the form
$$\omega ^r (d \omega )^{n-r-1}$$
where $\omega $ is some real valued form (see \cite{schw} for more
details), which is actually the case for the BHGV class, then the
corresponding elliptic complex is the de Rham complex, hence the result.\\

Some properties of the Ray-Singer torsion (denoted $T$ in the sequel) 
might be helpfull:\\

1. $T=0$ for even dim manifolds.\\
2. The torsion of a Cartesian product equals torsion of first factor
raised to the power equal to the Euler characteristic of second factor
(simply connected case).\\
3. For complex manifolds one has to use the Dolbeault complex. In this
case
for a Riemannian surface (complex dim 1) with genus 1 $T$ can be expressed
in terms of elliptic zeta functions, for higher genus one uses Selberg's
zeta function.\\
4. For the complex torus of complex dim greater than 1, one has that
$T=1$.\\
5. For Hopf surfaces (complex dim 2) one has that the torsion equals the
torsion of the torus which is its fibre over the sphere (for these results
and more see \cite{ray}).\\

The above discussion actually dictates that our foliation picture to
describe branes should rather correspond to a \emph{free}, i.e.
non-interacting theory. The natural generalisation then for interacting
5-branes would be to consider the full 11-dim Chern-Simons theory with
some non-abelian gauge group. This theory has been proved to be completely
soluble in dim 3 using \emph{geometric quantisation} methods. We know that
for large $k$, using stationary phase approximation (following terminology
of \cite{witten}), the theory becomes
abelian. The problem in dim 11 is considerably harder though. The main
point is that in dim 3 one can reduce the problen essentially to Riemann
surfaces by performing "surgeries" on the original 3-manifold. We do not
know for the moment if these \emph{surgeries} can be performed in dim 11.
Moreover, in contrast to dim 2 case, 
geometric quantisation for
10-manifolds is still at its infancy as a theory.\\

Another point to consider is the following: if one wants to get a
non-abelian generalisation as described above, one might loose S-duality.
This is an abelian symmetry, namely the isomorphism given by the Hodge
star operator for real valued forms. Although we have observed in
\cite{z1} that the Hodge isomorphism holds for \emph{flat} bundle valued
forms
(even with a non-abelian structure group), it brakes down in the general
case. Let us recall however that this is the on-shell case for
Chern-Simons theory (i.e. the Euler-Lagrange equations simply read that
the connection is flat).\\

Since the partition function of the BHGV class gives the Ray-Singer
analytic torsion, this is a topological invariant. Hence if we could find
an appropriate 11-form which could characterise foliations with codim 8,
namely to start with membranes insted of 5-branes, the result would be the
same. Hence S-duality holds. In other words it does not really matter
\emph{how} we foliate our manifold, clearly a manifold can be
foliated in many different ways and in many different codimensions, since
the partition function
of the characteristic class which describes the foliation is a topological
invariant. This guarantees S-duality.

The final comment would be that these "topological terms" in the
Lagrangians related to
characteristic classes for foliations (very closely related to non-linear
$\sigma $ models) are also met in some other interesting cases, namely
$\theta $ vacua and QCD, massive 3-dim Yang-Mills and 3-dim gravity as
well as fermions coupled to gauge fields again in dim 3 and the
non-conservation of parity (see \cite{red}, \cite{zee}, \cite{deser}). All
these cases refer to odd dim real manifolds. Maybe of some relevance also
for even (real) dim manifolds whose dim is an odd multiple of 2, using
complex structure and the Dolbeault complex.

\end{document}